\documentclass{PoS}

\title{Sensitivity of Type I X-Ray Bursts to \textsl{rp}-Process Reaction Rates}

\ShortTitle{Sensitivity of Type I X-Ray Bursts to \textsl{rp}-Process Reaction Rates}

\author{\speaker{A Matthew Amthor}$^{abcd}$, Daniel Galaviz$^{ab}$, Alexander Heger$^{bde}$,
	Alexander Sakharuk$^{ab}$, Hendrik Schatz$^{abc}$, Karl Smith$^{abc}$\\
\llap{$^a$}National Superconducting Cyclotron Laboratory, Michigan State University\\
	1 Cyclotron; MSU; East Lansing, Michigan 48824-1321; USA\\
\llap{$^b$}Joint Institute for Nuclear Astrophysics (JINA)\\
\llap{$^c$}Department of Physics and Astronomy, Michigan State University\\
	East Lansing, Michigan 48824-1116; USA\\
\llap{$^d$}Theoretical Astrophysics Division, Los Alamos National Laboratory\\
	MS B227; Los Alamos, New Mexico 87545; USA\\
\llap{$^e$}Department of Astronomy and Astrophysics, University of California, Santa Cruz\\
	Santa Cruz, California 95064; USA
}

\abstract{First steps have been taken in a more comprehensive study of the
dependence of observables in Type I X-ray bursts on uncertain
(p,$\gamma$) reaction rates along the \textsl{rp}-process path.  We use
the multi-zone hydrodynamics code \textsc{Kepler} which implicitly
couples a full nuclear reaction network of more than 1000 isotopes, as
needed, to follow structure and evolution of the X-ray burst layer and
its ashes.  This allows us to incorporate the full \textsl{rp}-process
network, including all relevant nuclear reactions, and individually
study changes in the X-ray burst light curves when modifying selected key
nuclear reaction rates.  In this work we considered all possible proton 
captures to nuclei with $10<Z<28$ and $N\le Z$.  When varying individual 
reaction rates within
a symmetric full width uncertainty of a factor of $10^4$, early results 
for some rates show changes in the burst light curve as large as 10 percent 
of peak luminosity.  This is very large compared to the current sensitivity 
of X-ray observations.  More precise reaction rates are therefore
needed to test current X-ray burst models, particularly of the burst rise, with
observational data and to constrain astrophysical parameters.
}

\FullConference{International Symposium on Nuclear Astrophysics - Nuclei in the Cosmos - IX\\
		 25-30 June 2006\\
		 CERN, Geneva, Switzerland}

\begin{document}

\section{Introduction}
Type I X-ray bursts are explosive events on the surface of accreting neutron stars.  They occur for 
certain accretion rates at which the hydrogen and helium nuclear fuel are largely consumed through the 
rapid proton capture process (\textsl{rp}-process) and the $\alpha$p-process, respectively. 
The \textsl{rp}-process 
takes place under conditions of extreme temperature and density in regions of proton excess 
and thus functions as a series of fast proton captures and weak decays involving isotopes on 
the proton-rich side of stability, including several near the proton drip-line.  The process 
is therefore sensitive to nuclear structure properties far from stability, mostly unknown 
to nuclear physics.  

Multi-zone burst models allow the study of the evolution of the neutron star crust composition 
over time, from accretion through stable burning, through the burst, and following the burst.  This 
allows consideration of effects which are impossible to study using only single-zone models.  
Through compositional inertia, the ashes of earlier bursts affect the character of later 
ones.  Also, the reheating of burst ashes by successive bursts will further process the nuclear 
material.  The resulting abundances in deeper layers of the neutron star crust, specifically 
that of carbon, are relevant to superburst processes.

This work examines the sensitivity of Type I X-ray bursts to all possible proton captures (those which produce particle stable compound systems) on nuclei with $10<Z<28$ and $N\le Z$, varying one rate at a time within an assumed uncertainty, with the goal of determining the extent to which our ability to model these systems is limited by our current knowledge of the nuclear physics inputs.  Similar studies conducted previously used postprocessing techniques~\cite{hix} or analytical methods~\cite{izz} in single-zone models, but because the \textsl{rp}-process and $\alpha$p-process to be studied are the main energy sources, such a treatment is inadequate here.  We used a specially adjusted, single-zone burst model to make a rough search for sensitivity in the burst light curve and nucleosynthesis in the full set of reactions.  These results were compared, in selected cases, to the results of more rigorous multi-zone simulations.  

\section{Reaction Rates}
Hundreds of uncertain proton capture rates lie on the \textsl{rp}-process path and can individually or collectively influence the energy generation and/or nucleosynthesis that will take place during a burst.  Direct measurement of these reaction rates is difficult~\cite{bis}.  Unknown compound reaction rates are frequently estimated using statistical model calculations.  These calculations, however, are only valid for capture to a region of sufficiently high level-density in the compound nucleus.  At burst temperatures this condition is often not satisfied for proton capture on lighter nuclei, $Z \lesssim 50$, near the proton drip line~\cite{rau}.  Proton capture rates to compound systems with low level density near and above the proton threshold will be dominated, at burst temperatures, by a few resonant capture contributions.  Therefore, the properties of these lowest energy proton-unbound levels critically determine the capture rates.

\begin{figure}[bt]
\begin{center}
\includegraphics[width=0.65\textwidth]{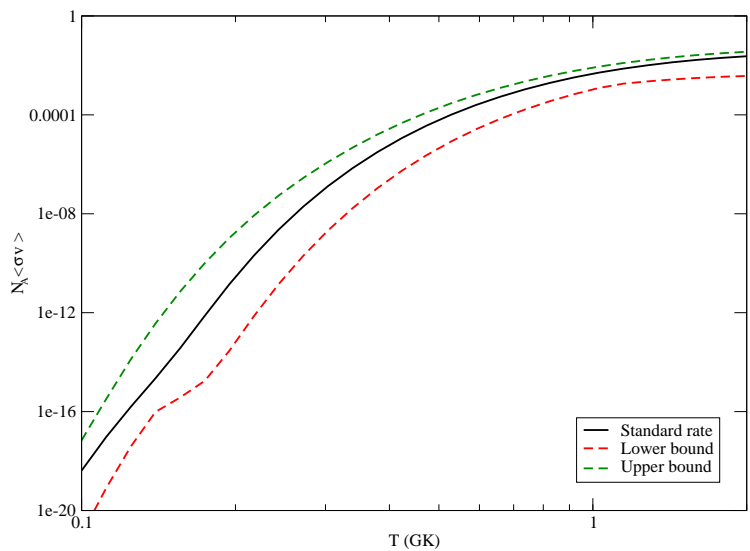}
\end{center}
\caption[ti42plot]{Shown is the calculated uncertainty for the $^{42}$Ti(p,$\gamma$)$^{43}$V reaction rate, assuming 100 keV uncertainty in resonance energies (Color online).}
\label{ti42plot}
\end{figure}

In the absence of experimental structure information on such isotopes along the rp-process path, shell model calculations are used, sometimes together with information from the mirror nucleus.  The commonly assumed 100 keV uncertainty in the excitation energy of theoretically calculated levels in these cases can yield orders of magnitude uncertainty in the reaction rates for temperatures relevant in Type I X-ray bursts.  For example, Figure~\ref{ti42plot} shows the rate and uncertainty for the $^{42}$Ti(p,$\gamma$)$^{43}$V reaction, calculated using structure information from Herndl et al.~\cite{her}.  These calculated uncertainties have been shown to be reasonable in recent experiments~\cite{cle,gal}. In the burst simulations of the present study, we assume that such a situation is typical and adopt a temperature-independent, symmetric uncertainty with a full-width of a factor of $10^{4}$.

\section{Single-zone model}
One-dimensional, multi-zone models are too computationally expensive to consider all possible rate uncertainties; so we begin with single-zone burst simulations.  The single-zone model of Schatz et al.~\cite{sch} was used, taking abundances and thermodynamic conditions from the burst zones in \textsc{Kepler}~\cite{woo} as initial conditions.  The specific zone and initial conditions used were adjusted in order to best reproduce the lightcurve and final abundances seen in the multi-zone model when using reaction rates unchanged from the current rate library.  It is perhaps worth noting that the reaction rate library being used by the single-zone model~\cite{sak} is not exactly the same as that used by \textsc{Kepler}~\cite{woo}.  

The process of running many single-zone simulations with changes to specific proton capture rates was automated, and each burst simulation took approximately 1.5 cpu minutes.  Very few of the initially considered rate variations caused a significant change in the burst lightcurve, but the changes that were observed were not subtle.  The changes seen in the burst ashes were generally less pronounced with a more continuous distribution of standard deviations.  In general, the lower bound reaction rates showed fewer and less significant changes from the baseline in both the lightcurves and the burst ashes than did the upper bound rates.  There is little initially apparent correlation between changes to lightcurves and to the ashes produced.

In cases where significant changes were observed, it was then more carefully considered to what extent the assumed uncertainty was justified for the specific reaction.  One of the most dramatic changes was seen when increasing the $^{26}$P(p,$\gamma$)$^{27}$S rate.  But since $^{27}$S has a proton separation energy, estimated by systematics, of $800$ keV and the lowest excited state in the mirror nucleus is at $1720$ keV, it is unlikely that the capture on $^{26}$P is dominated by uncertain resonant contributions.  This rate may still be a controlling factor, but should be varied over a smaller uncertainty.  Shell model calculations of the compound system could give us a clearer picture.  The rates for $^{46}$Cr(p,$\gamma$)$^{47}$Mn and $^{49}$Fe(p,$\gamma$)$^{50}$Co on the other hand, are likely dominated by resonant contributions from capture to poorly known levels in the compound nucleus, based on the structure of their respective mirror nuclei.  When increased, these rates both produce brighter bursts with shorter risetimes in the single-zone model.

In addition to individual rate variations considered here, higher-order effects should also be considered, as the sensitivity of the process to one rate may be strongly affected by the value of another uncertain rate or set of rates.  Naturally this leads to a much larger sample space which grows both exponentially with order and geometrically with the number of reactions considered.  Monte Carlo sampling methods, such as those used for novae in~\cite{hix}, could provide some valuable first insight into the broader set of these high-order dependencies.

\section{\textsc{Kepler} calculations}
The model used was that with solar metallicity and an accretion rate $\dot{M}=1.75\times 10^{-9} M_{\odot}yr^{-1}$, model \emph{ZM} as discussed in~\cite{woo}.  The first burst results from accretion to a bare iron surface and is therefore uncharacteristic of usually observed bursting behavior.  We look instead at the third burst, which, for standard reaction rates, shows properties very similar to those of the steady state behavior reached after a longer chain of bursts.  Of the several rates to which the single-zone model showed sensitivity and for which the assumed uncertainty was 
well justified, those considered so far in the multi-zone model include the 
upper limit rates for $^{42}$Ti(p,$\gamma$)$^{43}$V, $^{46}$Cr(p,$\gamma$)$^{47}$Mn, and 
$^{49}$Fe(p,$\gamma$)$^{50}$Co.  The lower limit rate for $^{46}$Cr(p,$\gamma$)$^{47}$Mn was also 
considered.  

\begin{figure}[tb]
\begin{center}
\includegraphics[width=0.95\textwidth]{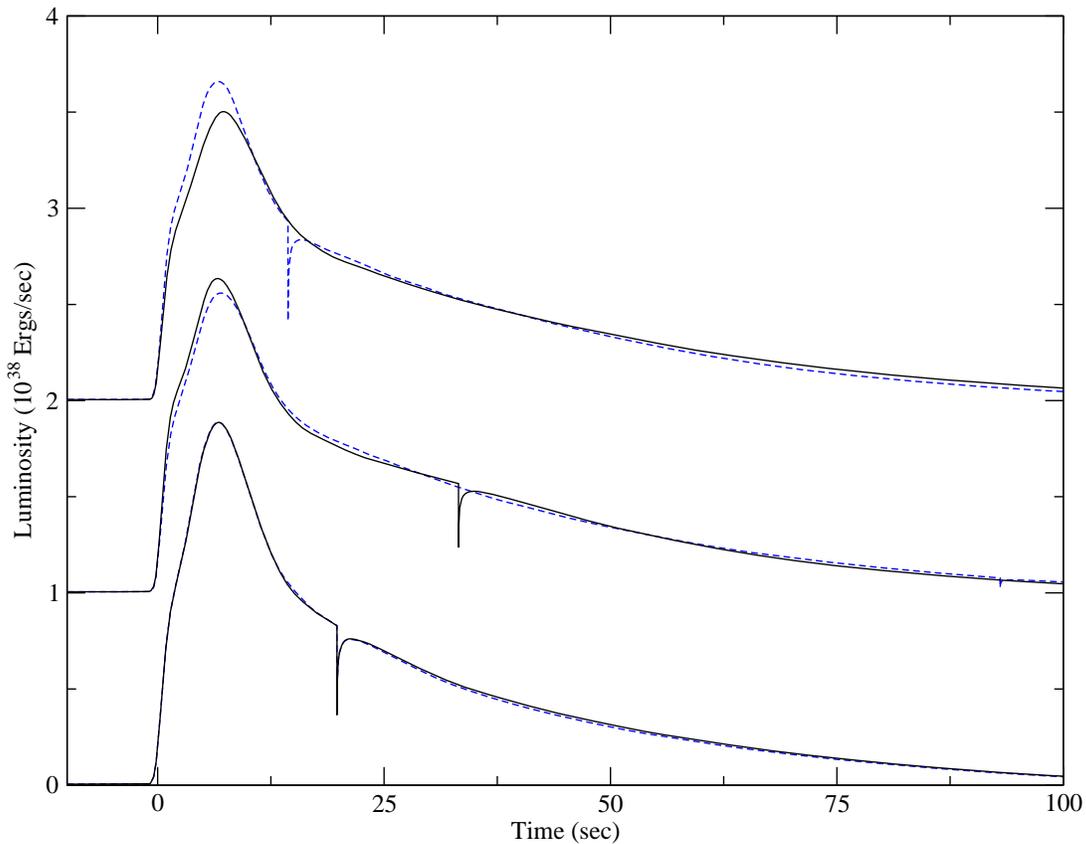}
\end{center}
\caption[cr46pulsesa]{Bursts one, two and three for the high metallicity, high accretion rate model in the \textsc{Kepler} code are shown, with later bursts shown successively higher on the graph.  Standard reaction rates yield the solid curves and an increased $^{46}$Cr(p,$\gamma$) rate yields the dashed curves.  In the third burst the increased rate results in an increase of over 10\% in peak burst luminosity.}
\label{cr46pulsesa}
\end{figure}

In the case of an increased $^{46}$Cr(p,$\gamma$) rate, we observe a similar large effect on the third burst in the multi-zone model to that seen in the single-zone model (see Figure~\ref{cr46pulsesa}).  For this particular reaction the single-zone 
model appears well calibrated.  This supports the conclusion that burst lightcurves are 
strongly dependent on this individual reaction rate.
In the case of $^{42}$Ti(p,$\gamma$), there are still slight differences in the decay of the lightcurve, 
but in the multi-zone model the increased rate results in faster cooling, the opposite of the effect seen in the 
single-zone model.  Also, the multi-zone simulations show faster rise and higher peak luminosity for this
upper-limit rate in the second and third burst, which were not seen in the single-zone model.
In the case of $^{49}$Fe(p,$\gamma$), the multi-zone model shows only a very slight increase in peak 
luminosity of the third burst, while the single-zone model showed an increase of over five percent in peak 
luminosity and a risetime $\approx 500$ ms faster.  

Clearly, a one zone model cannot be expected to correlate exactly with the results of a multi-zone simulation, especially during risetime, when convection plays an important role.  Nevertheless, this strategy can serve as an initial guide to help narrow the list of potentially important reaction rates.  It should be noted that, although the third burst seems fairly well converged, the studies of all rate sensitivities should be continued until a steady state behavior is reached, and ideally the exact same rate library should be used for both models to eliminate this as a possible source for discrepancies.

\section{Conclusions}

Based on the few reactions studied so far it seems that the one-zone approximation can help to identify 
critical rates in some cases.  While there can also be large differences in the sensitivities to rates 
between the two models, the one-zone approximation can be used as an initial guide to narrow down the 
list of potentially important rates to be studied with the multi-zone model.  This approach does not 
necessarily identify all rate sensitivities, but within the computing time limits it provides an 
effective way to determine some of them.  This observation further demonstrates the importance of considering effects which exist only in higher dimensions and will hopefully encourage work on codes that can consider Type I bursts in greater than one dimension.  In the future, the one-zone model analysis could be extended to the variation of multiple rates or to a full Monte Carlo analysis.  

There is at least one case found in this preliminary analysis with good evidence of a strong dependence 
of the burst lightcurve on a single reaction rate.  The other uncertain reaction rates considered in 
the multi-zone code to date appear not to be such critical inputs to the process for the specific accretion model considered.  

\section{Acknowledgments}
This work was carried out under the auspices of the National Nuclear Security Administration of the
U.S. Department of Energy at Los Alamos National Laboratory under Contract No. DE-AC52-06NA25396, and 
was also supported by the U.S. National Science Foundation Grants No. PHY-01-10253 (NSCL) and No. 
PHY-02-016783 (JINA).

\end{document}